\documentclass[apj, chicago]{emulateapj}

\journalinfo{\textsc{The Astrophysical Journal}}
\slugcomment{Received 29 November, 2018; Accepted 25 January 2019}
\usepackage{natbib}
\usepackage{epstopdf}
\usepackage[colorlinks,
        hypertexnames=false,
       pagebackref=false,
        linkcolor=red,      % color of internal links
    citecolor=blue,        % color of links to bibliography
    filecolor=magenta,   % color of file links
    urlcolor=red            % color of url
]{hyperref}                 %   package that links in pdf

\usepackage{array,etoolbox}
\preto\tabular{\setcounter{magicrownumbers}{0}}
\newcounter{magicrownumbers}

\newcommand{\rmd}{{\rm d }}
\newcommand{\nar}{ {\it NewAR}}

\shorttitle{Fast helicity injection leading to critically stable state and large eruptive activity in AR 12673}
\shortauthors{P. Vemareddy}
\begin{document}
\title{Very fast helicity injection leading to critically stable state and large eruptive activity in solar active region NOAA 12673 }
\author{P.~Vemareddy}
\affil{Indian Institute of Astrophysics, II Block, Koramangala, Bengaluru-560 034, India}
\email{vemareddy@iiap.res.in}

%%%%%%%%%%%%%%%%%%%%%%%%%
%% 							ABSTRACT 										
%%%%%%%%%%%%%%%%%%%%%%%%%
\begin{abstract}
Using the photospheric magnetic and coronal observations of Solar Dynamics Observatory, we studied the build-up and eruption of coronal non-potential magnetic structure in emerging active region (AR) 12673. The velocity field derived from tracked vector-magnetograms indicates persistent shear and converging motions of flux regions about the polarity inversion line (PIL). A major helicity injection occurs during rapid flux emergence consistent with the very fast flux emergence phase. While this helicity flux builds-up the sigmoid by September 4, the helicity injection by the continued shear and converging motions in the later evolution contributes to sigmoid sustenance and its core field twist as a manifestation of the flux rope which erupts after exceeding critical value of twist. Moreover, the total length of sheared PIL segments correlates with the non-neutralized current and maintains a higher value in both the polarity regions as a signature of eruptive capability of the AR according to the flux rope models. The modelled magnetic field qualitatively reproduces the sigmoidal structure capturing major features like twisted core flux as flux rope, and hook-shaped parts connecting at the middle of the PIL.  Study of quasi-separatrix-layers reveals that the sheared arcade, enclosing the flux rope, is stressed to a critically stable state and its coronal height becomes doubled from September 4-6. While demonstrating the fast injection of helicity per unit flux as the crucial factor for severe space-weather events, this study explains the formation of the flux rope and recurrent eruptive nature of the AR by the critically stable state of sheared arcade early on September 6. 
\end{abstract}

\keywords{Sun: coronal mass ejections (CMEs) ; Sun: evolution; Sun: fundamental parameters; Sun: magnetic fields; Sun: photosphere; Sun: activity }

%%%%%%%%%%%%%%%%%%%%%%%%%%%%
%
%%    INTRODUCTION 							%
%
%%%%%%%%%%%%%%%%%%%%%%%%%%%%
\section{Introduction}
\label{Intro}
%\linenumbers
The Sun often produces major eruptive phenomena that release vast amount of energy on the order of $10^{32}$ erg in a few seconds to minutes. It is now accepted that the source of energy for all of this solar activity is from magnetic fields. Active regions (ARs) are higher concentrations of magnetic field regions often seen with violent activity, like jets, flares, coronal mass ejections (CMEs), etc. The occurrence of these events relies on the information of triggering and driving mechanisms.  Even after large number of studies, not only the onset mechanism but also the supply mechanism of magnetic free energy is not yet clarified. Theoretically speaking, there are two effects that can supply magnetic free energy and magnetic helicity from below the solar surface to the corona. One is the so-called flux emergence activity, in which vertical motion carries magnetic fluxes through the photosphere. If the emerging flux has magnetic helicity, it must work as a helicity injection and may efficiently supply free energy into the solar corona. Another mechanism is photospheric shear motion, which supplies magnetic free energy as well as magnetic helicity into the solar corona by generating magnetic shear in the coronal field. Thus, triggering of CMEs, flares has mostly concentrated on the problem of evolution of a magnetic field in the very tenuous highly conducting plasma of solar corona \citep{,Forbes1995,linj2000,linj2003,forbes2006}.

Previous studies demonstrated various types of mechanisms that contribute dominantly to the accumulation of free magnetic energy in the solar atmosphere. These are 1) magnetic flux emergence or cancellation \citep{zhanghongqi1995,chenshibata2000,zhangh2001,Sterling2010}, (2) shearing motion \citep{ambastha1993,zhanghongqi1995,demoulin2002,vemareddy2012c,vemareddy2017b}  (3) sunspot rotation \citep{brown2003,tian2006,yanxl2008,vemareddy2012b,torok2013,vemareddy2016b}. The AR 9236 was reported to produce recurrent CMEs at an average time period of 10 hr \citep{gopalswamy2005} and the associated flares were not long decay events (LDEs). Report of the same AR by \citep{nitta2001} suggested the emerging magnetic flux as being responsible for repeated CMEs. Some studies also showed successive CMEs (e.g., AR 8038, 12371) in a timescale comparable to energy buildup by foot point motions. These were decaying ARs where the prolonged flux cancellation by converging motions and subsequent magnetic gradient increase about the polarity inversion line (PIL), introduce energy build-up in the AR magnetic system, which erupts into CMEs/flares \citep{shibu2000,liy2004,liy2010,vemareddy2017b,Vemareddy2018}. In the AR 12158, the two successive CME eruptions were triggered by helical-kink instability under the driving conditions of predominant sunspot rotation in a timescale of days \citep{vemareddy2016b}. Some ARs present with a combination of mechanisms in action. For example, recent reports on AR 11158 suggest that the shear and rotational motions of the observed fluxes played a significant role in transient activity with flares and CMEs \citep{sunx2012,vemareddy2012b}.  Based on these mechanisms, the eruptive scenario of ARs under a particular evolving condition of boundary motion have been numerically modelled \citep{antiochos1999,amari2003a,amari2010,archontis2014}. All of these models are based on the physical concept that the foot point motions predominantly contribute to a coronal helicity budget to form a twisted flux rope (FR) during or before its ejection as CME. 

The magnetic helicity describes the magnetic field complexity, including the twist, writhe, knot, and linkages of magnetic field. When coronal magnetic field is being pumped with helicity and energy, the magnetic complexity and non-potentiality increases. Conventionally, the magnetic complexity and non-potentiality are described in terms of parameters such as the magnetic shear \citep{Hagyard1986,WangT1994}, horizontal gradient of longitudinal magnetic field\citep{Falconer2003,SongH2006}, electric current \citep{leka1996,WangT1994}, twist parameter $\alpha$ \citep{pevtsov1994,hagino2004,Tiwari2009}, magnetic free energy \citep{Metcalf2005a}, current helicity \citep{Abramenko1996,ZhangBao1999}, etc. On the other hand, the EUV observations of the corona is seen with sigmoidal structures, a form of twisted flux rope associated with large-scale electric currents. From several such above studies of simultaneous magnetic field and coronal observations, it is now believed that the magnetic flux rope is built up by the line-tied photospheric motions, such as the magnetic flux emergence or the horizontal flows which injects the magnetic helicity into the higher solar atmosphere increasing the twist and kink of a flux rope (self-helicity) and the linkage between different flux ropes (mutual helicity). The magnetic helicity is conserved in an ideal MHD process and changes very slowly in a resistive process \citep{Taylor1974}. Thus, a flux rope with continuous injection of magnetic helicity inevitably erupts to remove the accumulated helicity, a manifestation of the CME. In this way, the magnetic helicity in the 3D volume is just a result of the helicity flux flowing into and out of the surface.  \citet{Guoy2013}, for the first time, found a quantitative relationship between the helicity injection and the twist number of the MFR. Thus a better understanding of the solar eruptions, on one hand, requires the unravelling the magnetic configuration for the flux rope, and the mechanism for the energy storage on the other hand. This kind of study helps in revealing the connections of the nature of boundary evolution and flux rope formation. 

From the magnetohydrodynamic (MHD) point of view, the flux rope is in equilibrium under the balance of magnetic pressure in the flux rope and the magnetic tension of the overlying magnetic field. If the twist number increases to some critical value, then kink instability would occur \citep{hood1979,torok2004}. On the other hand, if the decay index of the background field, in which the MFR is embedded, is larger than some critical value, torus instability can occur \citep{kliem2006,aulanier2010,demoulin2010}. \citet{demoulin2010} pointed out that the loss of equilibrium and the torus instability are two different views of the same physical mechanism, which is the Lorentz repulsion force of electric currents with different paths. In addition to the role play in the flux rope formation, the flux cancellation and tether cutting reconnection have also been invoked to account for the triggering the loss of equilibrium of the flux rope \citep{ballegooijen1989,moore2001,amari2003a,amari2003b}. 

Because lack of direct routine observations of the coronal magnetic field, the extrapolation of observed photospheric field by force-free field approximation is typically used for the study of the AR magnetic structure \citep{Wiegelmann2012}. The force-free model to the coronal field is justified by the low-$\beta$ plasma and higher (several hundred km/s) Alfven speed compared to photospheric flow speed. With this model, the coronal field evolution, corresponding to the slow photospheric plasma motion, is approximated as a quasi-static evolution of force-free equilibria. This enables one to study the buildup of pre-eruptive 3D structure, like a sigmoid or flux rope, then to find hints of the most appropriate configurations leading to eruptions (e.g., \citealt{savcheva2012a,xudong2012a,Guoy2013,jiangc2014,vemareddy2014a,vemareddy2016b,Vemareddy2018}).

The structure of the magnetic field is usually characterized by the topological analysis of quasi-separatrix layers (QSLs)  which denotes the places where the magnetic field line connectivity changes dramatically \citep{priest1995,demoulin1996,titov2002}. Two important QSL shapes are found to have relationship with the pre-eruptive configuration. The bald patch separatrix surface (BPSS) is separatrix surface with field lines touching the photosphere along PIL section where the transverse magnetic fields cross from negative to positive polarity as opposed to potential field case. The BPSS forms typically in cancelling and converging flux regions in the process of flux rope formation and has sigmoidal shape viewed from above. Hyperbolic flux tube (HFT) is another QSL structure with an X-line configuration that forms underneath the rising flux rope of BPSS topology. Alternatively, the HFT structure also forms above the flux rope or shared arcade facilitating breakout reconnection \citep{antiochos1999} in a quadrupolar magnetic configuration. In this case, the overlying flux contains a coronal null point which is possibly formed by a new flux emerging into an inverse pre-existing field \citep{WuST2005,Torok2009}. 

\begin{figure*}[!htp]
	\centering
	\includegraphics[width=.98\textwidth,clip=]{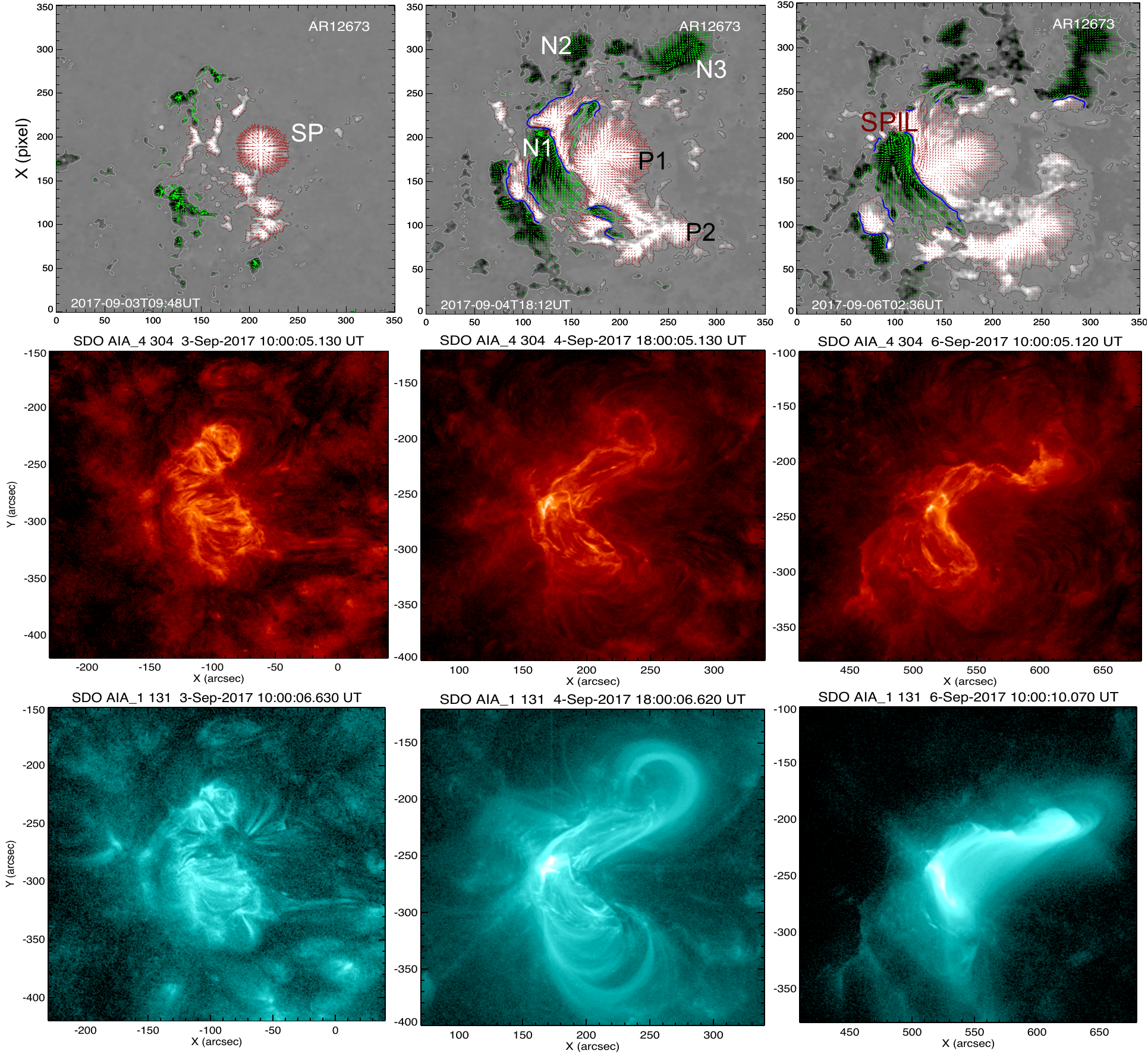}
	\caption{{\bf first row:} HMI vectormagnetic field observations of the AR 12673 on three different days. Background map is radial magnetic field with contours of $\pm$110G. Green/red arrows denote horizontal field vector proportional to magnitude. Blue curves are traces of SPILs with shear angle greater than 45$^\circ$. Axis units are pixels of 0.5arcsec units. {\bf second and third row:} Observations of the AR in AIA 304 \AA, 131 \AA~band.   }
	\label{Fig1}
\end{figure*}

In this manuscript, we studied the most violent AR 12673 producing strongest flares in the 24th solar cycle. Previous studies of this AR focused on qualitative energy supply mechanisms and triggering of major eruptions  \citep{YangS2017,VermaM2018,YanXL2018,HouYJ2018,LiuLijuan2018}. While the supply mechanism of energy and helicity is of prime importance, it is also very crucial to understand the rate at which they store in the corona to seek any relation to severe activity. This will be examined through a comparison with other ARs of various degree of activity in the form of strong flares/CMEs.  Further, in order to establish the connection of the helicity flux injection with the coronal field configuration like flux rope topology, we also studied the coronal field evolution by force-free extrapolation of the observed photospheric magnetic field. To this end, we estimate the key parameters like QSL, twist number, relative magnetic helicity, and total energy at different epochs of time evolution. In section~\ref{obs}, observations and overview of the AR is presented. Details of the results are given in Section~\ref{res}. Summary of the results with a possible discussion is given in Section~\ref{summ}.       

\begin{figure*}[!htp]
	\centering
	\includegraphics[width=.98\textwidth,clip=]{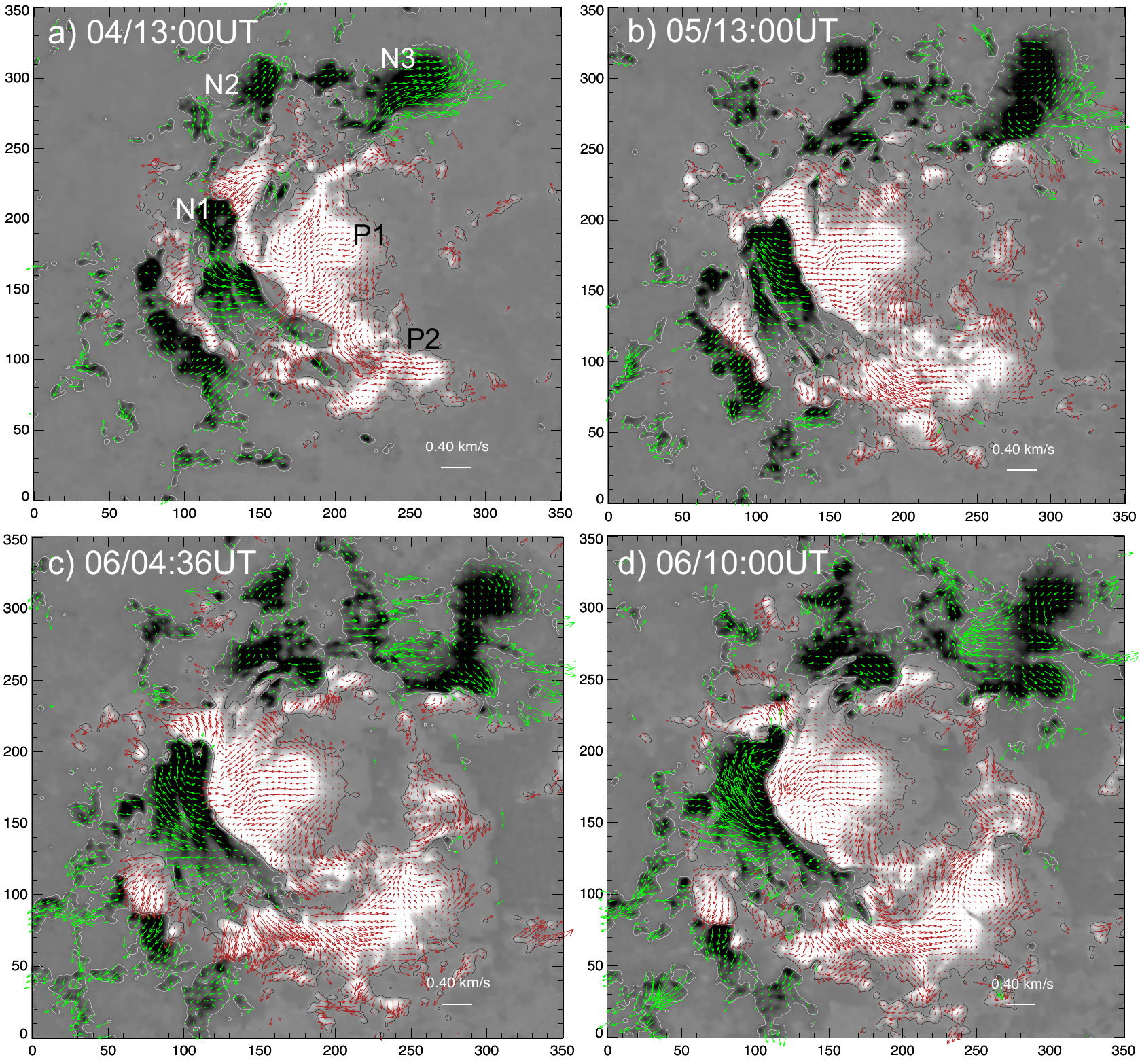}
	\caption{Velocity field of flux motions derived from DAVE4VM method. Background is HMI radial field map and the horizontal velocity is shown with red (green) arrows (normalized to 0.4km/s) in positive (negative) polarity.  The velocity patterns in N1 and P1 imply a predominant shear and converging motion of these polarity patches over the entire evolution time.  Axis units are pixels of 0\arcsec.5 units.}
	\label{Fig_vel}
\end{figure*}
\section{Observations and Overview}
\label{obs}
In this study, we use vector magnetic field observations at a 12 minute cadence and 0\arcsec.5 per pixel obtained from Helioseismic and Magnetic Imager (HMI; \citealt{schou2012}) aboard Solar Dynamics observatory (SDO; \citealt{pesnell2012}). Details of retrieving vector field from stokes vectors of filtergrams and other related information about HMI data products can be referred in \citet{hoeksema2014,bobra2014}. These disambiguated vector observations of the AR patch in the native coordinate system (latitude, longitude) are remapped to disk center by cylindrical equal area (CEA) projection method such that the AR patch center matches the disk center and are provided as \texttt{hmi.sharp\_cea\_720s} data product. The data product contains essentially $(B_z, -B_y, B_x)$ in the local heliographic coordinate system (see the appendix section in \citealt{sunx2013a}). Supporting coronal imaging observations are obtained from Atmospheric Imaging Assembly \citep{lemen2012} at a cadence of 12 s and 0\arcsec.6 per pixel.

In Figure~\ref{Fig1} (first row), the HMI vector magnetic field observations of AR 12673 are displayed. The corresponding coronal observations in AIA 304 \AA~, 131 \AA~are also shown in the second and third row panels. AR 12673 starts emerging from September 2 at disk position E11S08 into pre-existing positive sunspot polarity (SP). As described by \citet[][Figure 1]{YangS2017}, the emergence on September 3 is through two bipolar regions nearby a pre-existing sunspot of positive polarity.  During this time, the coronal images of AIA 304 \AA~, 131 \AA~pass bands indicate simple magnetic structure as seen in Figure~\ref{Fig1}. Following this on September 4, another two dipoles emerged within the existing patches. Notably, the first pair of the dipolar regions separated in east-west directions, whereas the later pair along north-south direction \citep{HouYJ2018}. The proper and shear motion of these emerging patches was interacted by the pre-existing spot and formed a large sheared main PIL. For convenience, we labelled the prominent polarity patches as P1, P2, N1, N2, N3 in Figure~\ref{Fig1}. These polarities are overally separated by a PIL of semi-circular shape. By September 4, the coronal images show an inverse-S sigmoidal structure (bent shape at the middle) with two hook shapes joining at the middle of main PIL between N1, P1.   

\section{Results}
\label{res}
\subsection{Magnetic evolution}
\subsubsection{Shearing and converging motion}
From the time series (every 12 minutes) vector magnetic field data, we derived the vector velocity field by using DAVE4VM \citep{schuck2008}. In Figure~\ref{Fig_vel}, we plot the horizontal velocity (arrows) on $B_z$ map. Contours of 100 G are also overlaid to identify a polarity patch boundary. Different features move with different velocity at different epochs, where the velocity is spread up to a maximum value of 0.8 km\,s$^{-1}$. For a large scale flow patter, we have averaged the velocity maps for over 2 hours (10 time shots), which then reduces the flow velocity to 0.4 km\,s$^{-1}$. The velocity field in N1 is coherent with a net organized flow pattern in northward, whereas that in P1 is southeast. Anti-clockwise whirl-pool motion pattern is also seen that is consistent with the sunspot rotation of N3, N1 and P2 which were reported to play major role in triggering two major eruptions on September 6, 2017  \citep{YanXL2018}. However, these sunspot rotations are smaller in spatial and time scales compared to large fraction of flux involved in shear motion about the mail PIL over entire evolution period. Therefore, these  motions indicate a predominant shearing and converging motion of N1 and P1 over the entire evolution time and is suggested to play prime role in energy and helicity storage in the magnetic system. Consequently, the sigmoidal structure builds as seen in corona after a day of the AR emergence and is persisted for days through the continuous helicity and energy input by these motions as earlier studied cases for example \citet{vemareddy2017b}. These observations are consistent with \citep{VermaM2018} linking the shear flows and head-on collision of new and pre-existing flux with the origin of two X-class flares.

\subsubsection{Net magnetic flux}
In Figure~\ref{Fig_met}, we plot evolution of magnetic parameters related to build-up of non-potential nature of the AR. A rapid emergence commenced early on September 3 growing sunspot groups and forming full AR in a time of a day. In Figure~\ref{Fig_met}(a), net flux in positive ($B_z>0$) and negative ($B_z<0$) polarities are plotted with time. Disk integrated GOES X-ray flux is also shown in the same panel with y-axis scale on right side. From September 4 to September 10, the AR produced a total 4 X-class flares, 27 M-class flares and a multiple of small flares \citep{YangS2017}. From this GOES X-ray light curve, we can divide the flaring activity into two phases as indicated by orange shades. The first phase starts from early September 4 and ends on September 5 at around 18:00\,UT. This phase correlates with the flux emergence and includes M-class flares. The second phase starts from 11:00\,UT on September 6 and continues till September 10. This phase is the most energetic with the largest X-class flares in solar cycle 24. The net flux varies 2--10$\times10^{21}$ Mx in each polarity on September 3rd, due to rapid emergence of flux. The later evolution follows gradual flux emergence and its areal spreading, increasing the net flux to $34\times10^{21}$ Mx till end of September 8. As found in preliminary study of \citet{SunX2017}, the AR has the fastest flux emergence of any observed values at an average of $4.93\times10^{20}$ Mx\,hr$^{-1}$ over 5-day period. 

\subsubsection{Net electric current and SPILs}
While growing, the interaction of opposite polarity regions creates compact regions forming sheared polarity inversion lines (SPILs) at their interface. The opposite motion of polarity regions parallel to PIL is referred to shear motion and generates stress in magnetic field connecting those regions. In that case, the adjacent field vectors are parallel to PIL, and the extent of the shear is measured by $\theta = cos^{-1} (\mathbf{B}_o \cdot \mathbf{B}_p / |\mathbf{B}_o| |\mathbf{B}_p|)$, \citep{ambastha1993} where $\mathbf{B}_o$ is the observed field and  $\mathbf{B}_p$ is the potential field. Thus, sheared PIL (SPIL) is a measure of stressed magnetic configuration in the AR.  In Figure~\ref{Fig1}, the HMI vector magnetic field observations of AR 12673 are displayed. We also traced the sheared PILs by an automated procedure similar to the one applied for tracing PILs of strong vertical field gradients \citep{mason2010} and applied in a statistical study \citep{Vasantharaju2018}. In this procedure, we smooth the $B_z$ map to a smoothing factor of 8 pixels (4 arcsecs) and identified the zero Gauss contour with shear angle greater than 45$^\circ$.  The value of SPIL length is slightly dependent on degree of smoothness. The error in the SPIL length is estimated by varying smoothing factor, and could be upto 3-5\,Mm. The maps are overplotted with these traced SPIL segments. It is indicated that with the emergence of the AR, the interaction of opposite polarities increased the interface of SPILs between P1 and N1.

\begin{figure*}[!htp]
	\centering
	\includegraphics[width=.87\textwidth,clip=]{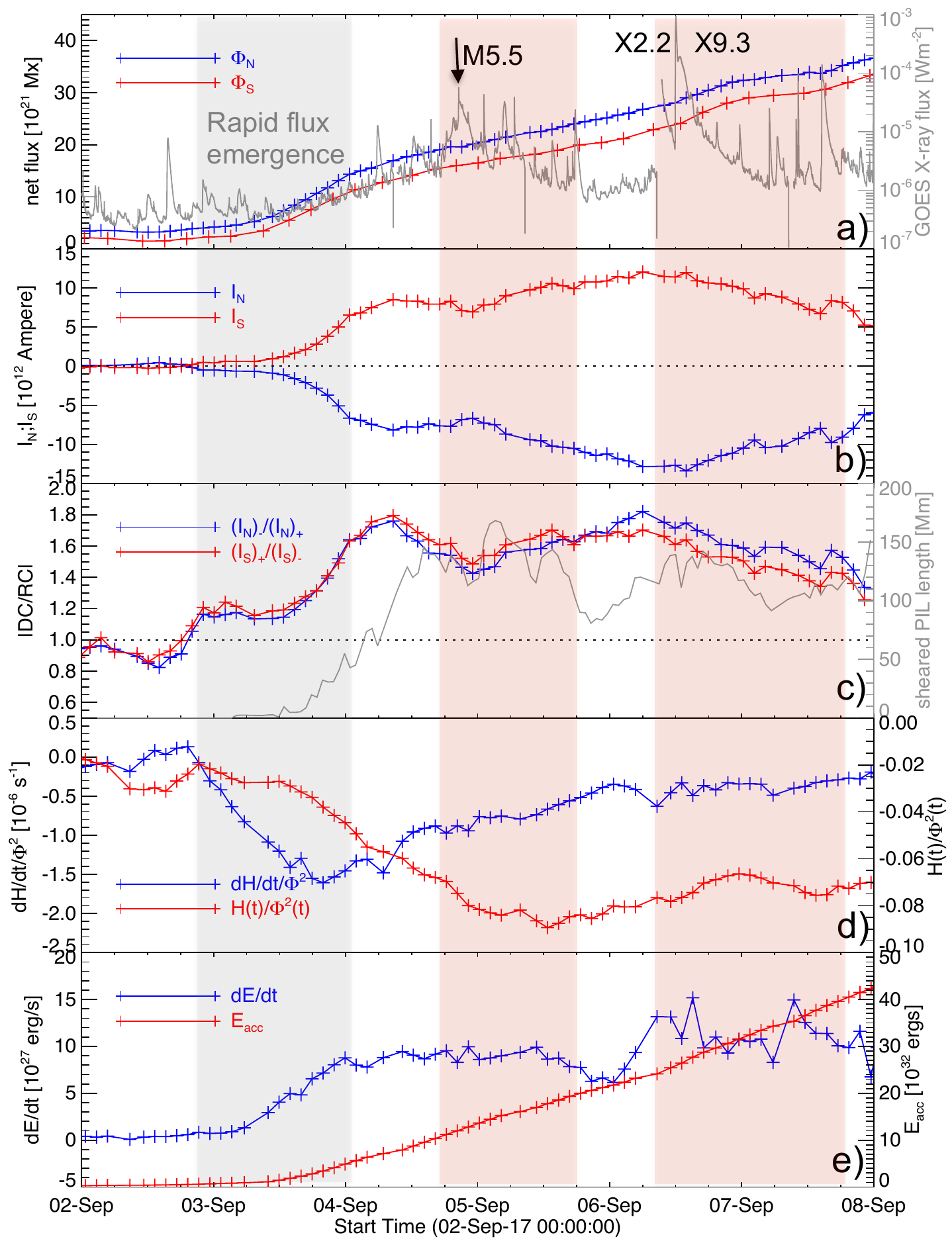}
	\caption{Magnetic evolution in AR 12673 a) net flux in positive (north) and negative (south) magnetic polarities. Disk integrated GOES X-ray (1.0--8.0 \AA~passband) flux is also shown with y-axis scale on the right, b) Systematic evolution of net vertical current from north and south polarities c) Neutralization of net current in individual polarities. Horizontal dashed line marks the neutralization level of unity. Total length of all SPIL segments are also plotted with y-axis scale on right. It follows the $|DC/RC|$ profile indicating the relevance of SPILs and degree of neutrality. d)time-rate of helicity flux normalized by averaged net flux of positive and negative polarities. Normalized accumulated helicity (time integrated helicity flux rate) is also plotted with y-scale on right. The normalized helicity flux reaches to 0.09 turns indicating the moderately twisted flux system. e) Energy flux injection and its accumulated quantity. Rapid flux emergence phase is marked with grey shade and the two major flaring phases are indicated with orange shade.  }
	\label{Fig_met}
\end{figure*}

Vertical component of electric current ${{J}_{z}}=\frac{1}{{{\mu }_{0}}}{{\left( \nabla \times \mathbf{B} \right)}_{z}}$ is another non-potential measure accounting horizontal field gradings and is readily computed with vector field observations. In Figure~\ref{Fig_met}(b), the net current ($I=\sum\limits_{N}{{{\left( {{J}_{z}} \right)}_{i}}} dA$, where dA is area of the pixel) obtained in each polarity is plotted with time. A threshold of $|\mathbf{B}|>150$ G is used for the reliability of the values above sensitivity, noise, inversion errors. The I increases with the emergence of the AR on September 3, and reaches to $5\times10^{12}$ A in magnitude in each polarity by the end of that day. The later evolution follows its further increase reaching a maximum value of $13\times10^{12}$ A around 12:00\,UT on September 6, when a major eruption with X9.2 flare occurred. Note that the I is negative (positive) in positive (negative) polarity, indicating a dominant negative chirality of the AR magnetic structure according to the definition of current helicity.  

In a magnetic polarity, the net current is theorized to be neutralized by canceling volume and sheath currents of the flux tubes \citep{parker1996}, which is found to nearly valid in isolated sunspot ARs \citep{venkat2009}. However, when interacting opposite polarities with SPILs exists, the net current breaks neutralization \citep{georgoulis2012}. According to the flux rope models of CMEs eruptions \citep{zakharov1986}, the breakdown of net current neutralization refers to a form of Lorentz force development and stability loss. As a reason, breakdown of neutrality is proposed to be a proxy assessing the ability of ARs to produce major eruptions \citep{YangLiu2017}. Following this, breakdown of net current neutralization is found to be correlated with the presence of SPILs and the observed activity in many ARs \citep{Vemareddy2019}. 

In Figure~\ref{Fig_met}(c), the ratio of direct current (DC) and return current (RC), as the dominant and non-dominant currents, in each polarity is plotted with time. HMI provides inversion errors of field vectors which vary upto 50 G. On considering an average error  of $\delta B_x=\delta B_y=40$ G in a typical distribution of $n=10^4$ pixels, we found that the range of uncertainty ($\sqrt{n}\times \delta J_z dx dx$) of a signed net current in a given polarity can never be larger than $0.1\times10^{12}$A \citep{vemareddy2017d, Vemareddy2019} . Here $dx$ is HMI pixel size of $0\arcsec.5$ Therefore, our estimation of $|DC/RC|$ can have a maximum error limit of 0.14, and is very small compared to the range of $|DC/RC|$ evolution in CME and flare producing ARs. 

\begin{figure*}[!htp]
	\centering
	\includegraphics[width=.99\textwidth,clip=]{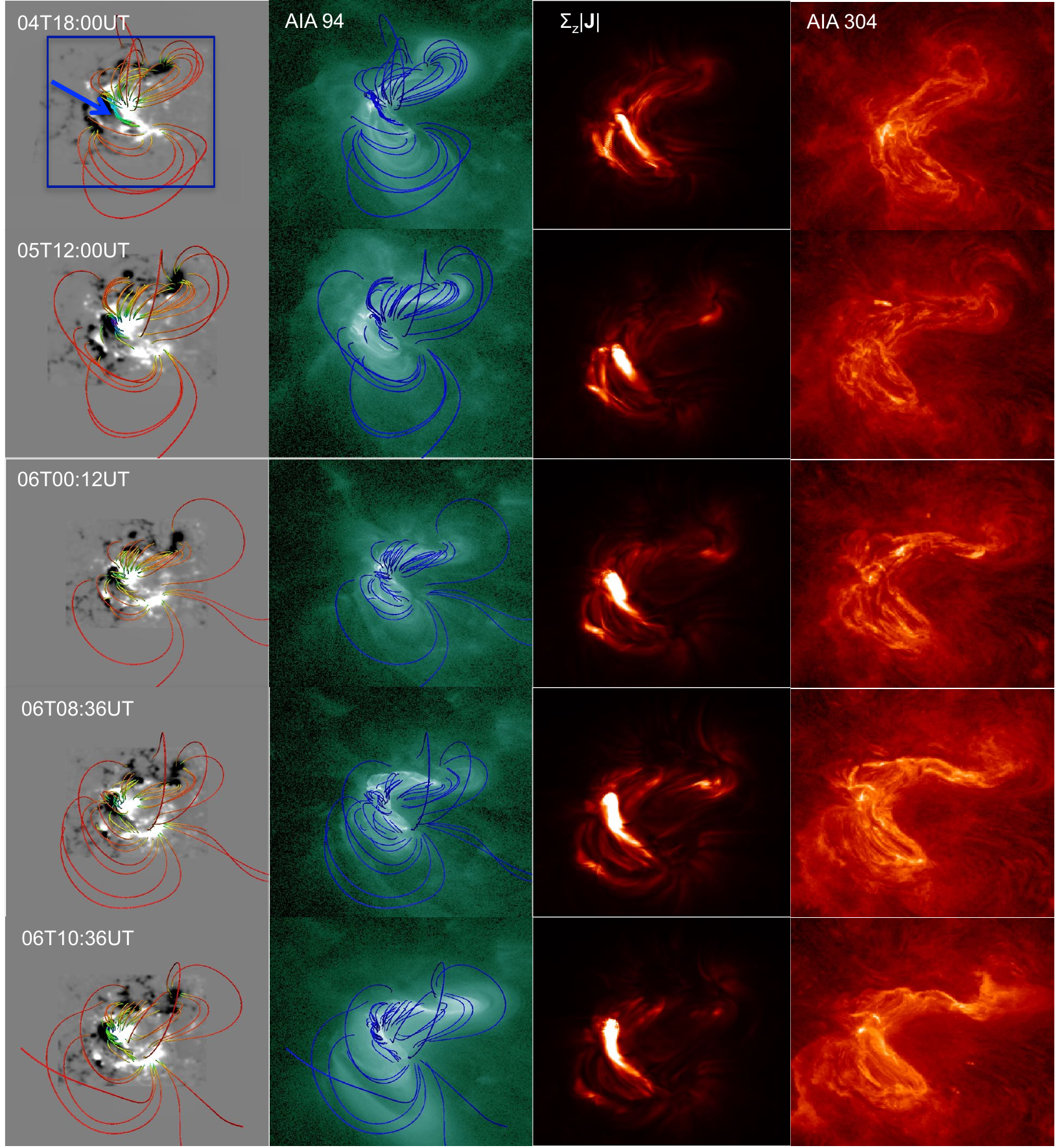}
	\caption{ Magnetic structure modelled by NLFFF {\bf first column:} magnetic field lines rendered on $B_z$ map at different epochs of AR evolution. Field line color is due to $|{\bf J}|$ distribution. Arrow indicates flux rope structure in 04T18:00\,UT panel. {\bf second column:} the same field lines on AIA 94\AA~channel images obtained at respective times. The global magnetic structure in the AR convincingly mimics the morphology of the plasma emission. {\bf third column:} vertically integrated electric current distribution ($\sum_z |\mathbf J|$) approximately depicting the plasma emission in AIA 94\AA~passband  shown in second column panels. The maps are scaled within $0-500$ Am$^{-2}$. Field-of-view is the region enclosed by rectangular box shown in 04T18:00\,UT panel of first column. {\bf fourth column:} AIA 304 \AA~observations showing the sigmoidal morphology. These panels indicates the plasma emission is related to dense current distribution produced by twisted magnetic structure in the corona.}
	\label{Fig_ext_j2d}
\end{figure*}

A value of $|DC/RC|$ at one indicates perfect neutralization and a greater value refers to degree of non-neutrality \citep{Torok2014}.  We can notice that neutrality breaks down with the emergence of the flux from 3rd September onward and reached to 1.4 by the end of the day. After the fast emergence phase, the $|DC/RC|$ varies about 1.5 in the later phase of evolution till 8th September. Importantly, the $|DC/RC|$ follows the total length of all SPIL segments plotted in the same panel with y-axis scale on right. This also substantiates for the relevance of SPIL amid the compact regions in the ARs and the breakdown of net current neutrality \citep{georgoulis2012}. Breakdown mainly occurs in the rapid emergence phase, restoring to neutrality in the later phase during which the flux regions separate and become isolated without SPILs. Unlike the case in emerging ARs \citep{Vemareddy2019}, the persistent higher degree of non-neutrality of net electric current throughout the evolution is an indicator of eruptive capability of this AR. 

\subsubsection{Pumping of magnetic helicity and energy}
While the AR flux emerges and spreads through plasma motions, the rate of build up of complexity is measured by the helicity flux injection through the photospheric surface of the AR given by \citep{Berger1984}
\begin{equation}
{{\left. \frac{dH}{dt} \right|}_{S}}=2\int\limits_{S}{\left( {{\mathbf{A}}_{P}}\bullet {{\mathbf{B}}_{t}} \right){{\text{V}}_{\bot n}}dS}-2\int\limits_{S}{\left( {{\mathbf{A}}_{P}}\bullet {{\mathbf{V}}_{\bot t}} \right){{\text{B}}_{n}}dS}
\end{equation}

where $\mathbf{A}_p$ is the vector potential of the potential field $\mathbf{B}_p$, $\mathbf{B}_t$ and $B_n$ denote the tangential and normal magnetic fields, and $\mathbf{V}_{\perp t}$ and $\mathbf{V}_{\perp n}$ are the tangential and normal components of velocity $\mathbf{V}_\perp$, the velocity perpendicular to the magnetic field lines. From the time series (every 12 minutes) vector magnetic field data, we derived the vector velocity field by using DAVE4VM \citep{schuck2008} and compute $dH/dt$ (see also \citealt{LiuYang2012,vemareddy2015a}). Although there is no way that we can actually compute errors, a Monte Carlo experiment is used to represent probable error in helicity flux computation. Here we randomly added noise of magnitude 100 Gauss to three components of the vector magnetic field, and repeated the vector velocity and helicity flux computations for 200 times \citep{LiuYang2012}. The maximum error is 1$\sigma$ error of all 200 experiments and was found to be 23\%. To represent the average complexity per flux tube (twist rate), we normalized the $dH/dt$ with the half of the unsigned flux $\Phi=\frac{1}{2}\int|B_z(x, y, z=0)| dxdy$  and plotted in Figure~\ref{Fig_met}(d). The fast emergence phase on September 3rd is accompanied by higher twist injection upto $-0.5\times10^{-6}$ turns/s. Note that the negative sign denotes the negative chirality of the AR. It refers that the emerging flux is having pre-existing twist from convection zone. This phase follows a decrease of twist injection rate to $-0.7\times10^{-6}$ turns/s where horizontal motions dominate over the gradual flux emergence over days. The time integrated $dH/dt/\Phi^2$ is coronal accumulation of helicity flux $H(t)/\Phi^2$  and is plotted in the same panel with y-axis scale on the right. The $H(t)/\Phi^2$  reaches to 0.09 turns by 5th September, indicating a moderately twisted flux system. We point that the helicity flux calculations in emerging ARs are more useful for the studies of CME occurrence because the values represent the AR flux without missing the pre-existing/emerged structure. Further, \citet{YanXL2018} interpret that the counter-clockwise rotating sunspots (here N1, P2, N3) inject negative helicity and relate with the successive X-class flares on September 6, 2017. Our results suggest that the AR is already in critically non-potential state by predominant shear motions over entire evolution period, which in addition of sunspot rotation could trigger eruptions on the September 6.  

The recurrent CME producing AR 12371 is found to accumulate 0.15 turns, whereas CME-poor flare-rich AR 12192 is having 0.02 turns \citep{vemareddy2017b} over a similar time of evolution. In the former case flux rope (sigmoid) structure is observed whereas no flux rope in the later. As suggested by \citet{vemareddy2017b}, a higher value $H(t)/\Phi^2$ denotes more twisted flux system like flux rope and have lesser confining flux. From these cases, we suggest that the magnetic flux normalized helicity flux is an important parameter in distinguishing strong erupting ARs.       
\begin{figure*}[!htp]
	\centering
	\includegraphics[width=.95\textwidth,clip=]{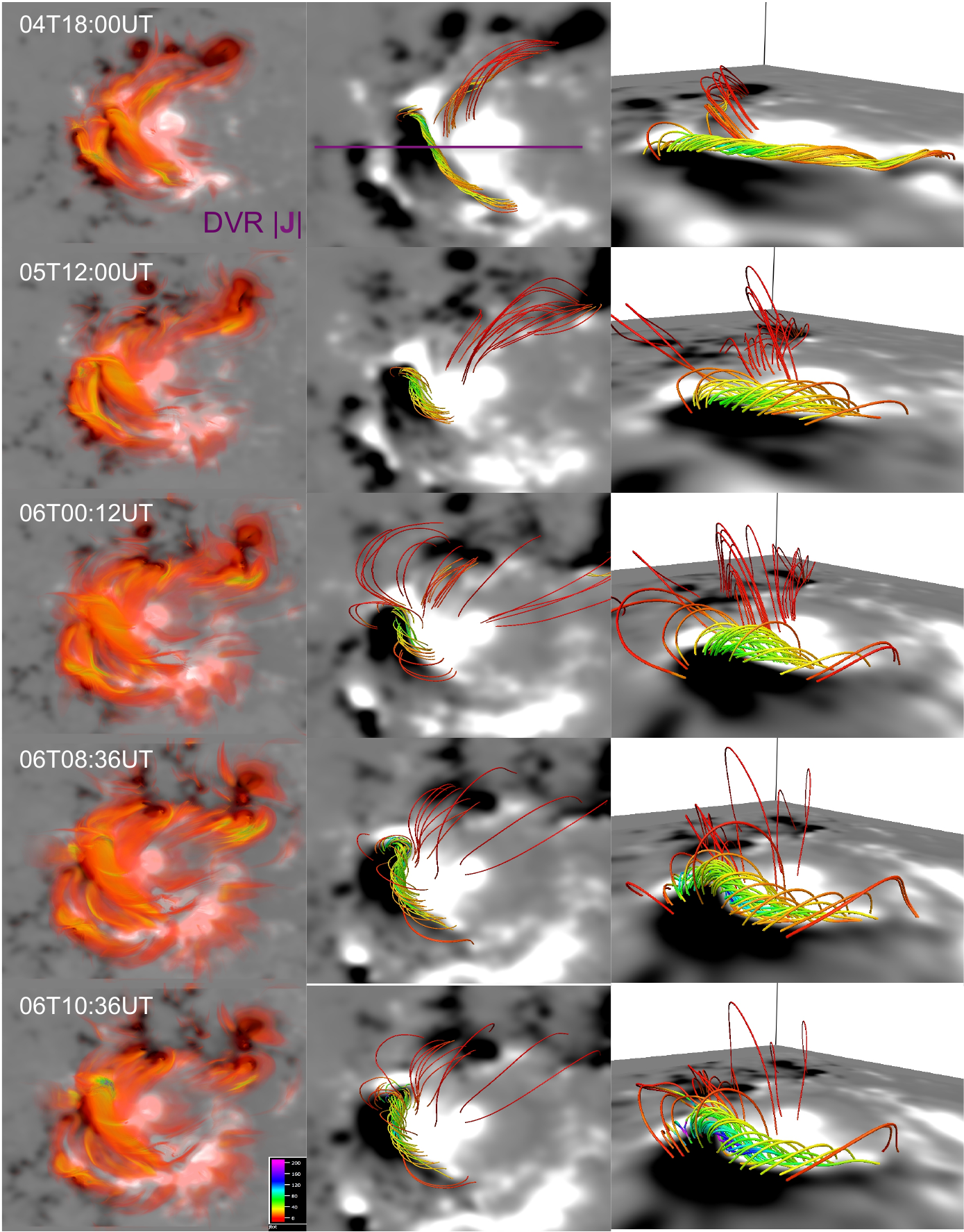}
	\caption{  NLFFF core field structure on different days of evolution in AR 12673. {\bf first column:} volume rendering of 3D distribution of $|\mathbf{J}|$. {\bf second column:} top view. A vertical slice is placed along horizontal magenta line for further analysis. {\bf third column:} perspective view. In all the panels, the structure mimics the twisted flux rope. The field lines rendered  are biased with intense $|\mathbf{J}|$ and are color coded accordingly.   	
	}	\label{Fig_core_fld}
\end{figure*}

Similarly, the energy flux injection (Poynting flux) across the surface \citep{KusanoK2002} 
\begin{equation}
{{\left. \frac{dE}{dt} \right|}_{S}}=\frac{1}{4\pi }\int\limits_{S}{B_{t}^{2}{{V}_{\bot n}}dS-\frac{1}{4\pi }}\int\limits_{S}{\left( {{\mathbf{B}}_{t}}\bullet {{\mathbf{V}}_{\bot t}} \right){{B}_{n}}dS}
\label{eq_dedt}
\end{equation}
is computed and plotted in Figure~\ref{Fig_met}(e).  Corresponding to the dH/dt profile, the energy flux injection $dE/dt$ (a positive definite quantity always) also shows a higher rate of input during rapid emerging phase which reaches to $8\times10^{27}$erg/s. Assuming this value of flux for two days, the coronal energy budget would be $1.35\times10^{33}$ergs that probably supplied for the M-class flaring activity on September 5. Following this phase, the $dH/dt$ maintains its constant influx at an average value of $7\times10^{27}$erg/s. Given this constant supply of energy flux, the accumulated quantity is $4\times10^{33}$ergs by the end of September 7. Notedly, $dE/dt$ in this AR is strong by a factor of two compared to less intense-flaring ARs \citep{vemareddy2015a,vemareddy2017b}. From this energy flux study, we suggest that the coronal magnetic field is constantly driven to a stressed state of critical energy level that is significant enough to power the sequential X-flares with CMEs.

\subsection{NLFFF model of AR magnetic structure}
The AR magnetic structure is reconstructed by performing nonlinear force-free field (NLFFF) extrapolation of the observed photospheric vector magnetic field \citep{wiegelmann2010}. In order to weaken the effects of the lateral boundaries, the observed boundary is inserted in an extended field of view and computations are performed on a uniformly spaced computational grid of $800 \times 800 \times 400$ representing physical dimensions of $291 \times 291 \times 146$ Mm$^3$.  At different epochs of AR evolution, the flux imbalance is less than 10\%. To satisfy the force-free conditions, the magnetic field components are pre-processed \citep{wiegelmann2006}. We first initiated the NLFFF code with potential field (PF) but found that the modelled field failed to reproduce the structured AIA emission especially the hook shapes. This is due to the fact that the large scale structure is not close to potential field. Therefore, the NLFFF code is initiated with the 3D linear force-free field (LFFF) constructed from the vertical field component of the observed field \citep{gary1989}. The force-free parameter used is obtained by minimizing the least-square difference of the modelled and observed transverse field and is known as $\alpha_{\rm best}$ \citep{pevtsov1994,hagino2004}. Direct use of this LFFF with $\alpha_{\rm best}$ parameter as initial condition for NLFFF results in over injection of magnetic twist into the field lines leading to their unexplained shape. We make several runs with varying $\alpha_{\rm best}$ and compare the final NLFFF structure with coronal images and then found that a value of half of the estimated $\alpha_{\rm best}$ would result in for the suitable initial model for NLFFF. The NLFFF relaxation is proceeded by minimizing the functional $L$ containing volume integral terms of Lorentz-force, magnetic field divergence and a surface integral term that accounts the measurement errors while injecting the boundary observations. The final solution is assessed by the current-weighted sine angle $sin\theta_J$ between magnetic field $\mathbf{B}$ and electric current density $\mathbf{J}$ and magnetic field divergence in the computational box \citep{Wheatland2000}.  In our cases of NLFFF extrapolation, the relaxation converges to  $10^{-3}$ of initial $L$ with 10-12$^\circ$ of $\theta_J$ and an average magnetic field divergence of the order of $10^{-4}$. Note that the initial condition of LFFF is implemented to reproduce the unexplained structure by NLFFF resulted with PF, and is constrained by the observed configuration of the coronal images after several trial runs with varying $\alpha_{\rm best}$.

\begin{figure}[!htp]
	\centering
	\includegraphics[width=.5\textwidth,clip=]{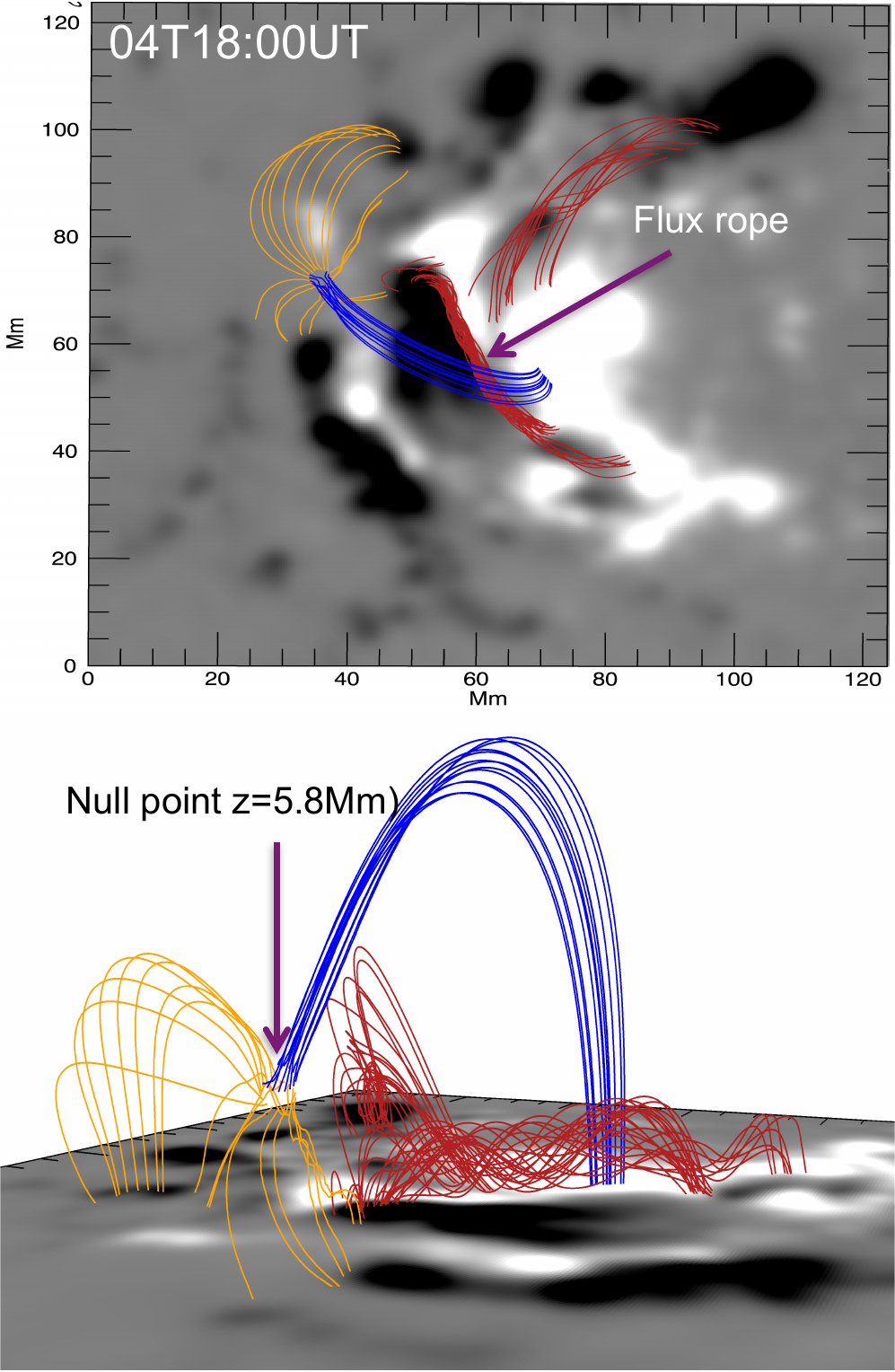}
	\caption{  Null point topology at 04T18:00 UT time frame. {\bf top:} top view. Fan field lines (orange) diverge from null point to negative polarities surrounding the positive polarity. Spine field lines (blue) overly the flux rope (red) with footpoints in the P1 and P2 polarities, {\bf bottom:} perspective view. Background image is $B_z$ distribution.       	
	}	\label{Fig_null_fr}
\end{figure}

The NLFFF magnetic structure at different epochs is shown in Figure~\ref{Fig_ext_j2d}. The rendered field lines are overlaid on $B_z$ map in the first column and on AIA 94 \AA~passband images in the second column panels. The field lines connecting P2, N1 form lower lobe and that connecting N3, P1 form another northern hook structure. The field lines anchored near the SPIL are inverse S-shaped and graze the PIL manifesting low lying flux rope (blue arrow in 04T18:00\,UT panel) and are overlaid by high lying potential like field lines. This flux rope structure persisted in all panels as indicated by arrow. This modelled structure mimics the sigmoidal shape very well exhibited in AIA 94 \AA~images. Especially the hooked structure in the top and bottom of the inverse S-sigmoid matched well. The global magnetic structure seems not changed much from September 4 to 6, which means that the photospheric flux motions maintains the global non-potentiality by sigmoidal shape. While there are active events releasing stored energy by the field reconfiguration, it is replenished by underlying photospheric shear and converging flux motions through quasi-static evolution. 

It is important to note the projection effect with observations. When the AR moves away from the disk center, the radial direction becomes increasingly departed from the line of sight, so in order to compare the modeled field lines with the coronal AIA observations, the best practice is to tilt the modelled magnetic structure by an angle of the AR position on the disk \citep{Guoy2016,Guoy2017a}. Although should have corrected, this projection effect has little contribution in our overlaid panels of AIA 94\AA~. Especially, the field lines in the lower limb of the sigmoid on September 6, which are higher in height, appear to deviate a bit from the emission pattern. 

In the third column panels, we display the vertically integrated electric current ($\sum_z |\mathbf{J}|$) distribution. The coronal field is driven by photospheric shear motions which naturally builds coronal volume currents in the stressed configuration. Corresponding to the sheared/twisted field structure (flux rope), the intense $|\mathbf{J}|$ are present along main PIL. The overall morphology of this current distribution is similar to the sigmoid observed in EUV 304 \AA~images displayed in fourth column panels and demonstrates the NLFFF model captures the most of the observed features.

In Figure~\ref{Fig_core_fld}, the rendered core field structure is displayed in top view (second column panels) and perspective view (third column panels). For a better representation of volume current with the field lines, we also display the volume rendering of $|\mathbf{J}|$ in the first column panels. In all panels, the background image is $B_z$ distribution. The field lines are selected at intense locations of $|\mathbf{J}|$ and are color coded accordingly. In all the panels on different days, the structure mimics the twisted flux rope above the main PIL. Especially, the NLFFF structure at 18:00 UT on September 4 is compact with coherent continuous twisted field lines. On other days, there exists highly sheared arcade in addition to low lying continuous inverse-S ones. However, the field lines connecting north-west bipolarities are part of sheared arcade in the top hook structure of the sigmoid. 

A careful examination reveals a null point topology in the 04T18:00 UT frame as depicted in the Figure~\ref{Fig_null_fr}. The null point position is located by algorithm described in \citep{vemareddy2014a}. Basically it involves scanning for the null to locate a possible grid cell and then finding the precise position using a tri-linear interpolation with the help of an iterative Newton–Raphson scheme within the grid cell. The null point is located above a positive polarity on the east side of the N1 polarity. It is at a height of around 5.8 Mm. Null point topology is typically associated with a fan-spine field line structure. The null point properties are described by eigen values and eigen vectors of the Jacobian $\delta B=\nabla_j B_i=\partial B_i/\partial x_j$ obtained in the vicinity of the null \citep{lau1990}. Two of the eigenvectors (with the same sign for the eigenvalues) define the fan surface and the third one specifies the spine direction. Knowing fan plane orientation by two of the eigen vectors, field lines away from the null can be traced in a circle of points on either side of the fan-plane to visualize the local null topology. Here the spine field lines overly the flux rope and have foot points spreading in P1 and P2 polarities as confining structure. This is one of the the proposed structures in a complex quadrapolar magnetic configuration facilitating breakout reconnection the underlying flux rope eruption \citep{antiochos1999}. Note that here the flux rope is under the spine field lines instead of fan field lines as for example in \citet{Guoy2016}. These kind of structures form in regions of new emerging flux into a preexisting flux of inverse configuration \citep{WuST2005,Torok2009}.  All together, the modelled field reproduces the complex magnetic structure resembling the observed EUV sigmoid with embedded twisted flux rope. 

\subsection{Quantitative estimates of energy, helicity, twist number}
The total magnetic energy is estimated by $\int_VB^2/8\pi dV$ in the AR volume. Owing to the net flux increase in the AR, the potential energy $E_p$ increases from $12.4\times 10^{32}$ ergs on 04T18:00 UT panel to $18.7\times10^{32}$ ergs at 06T11:24 UT panel. The total energy $E$ have increasing nature from $18\times10^{32}$ ergs on 04T18:00 UT to $26.6\times10^{32}$ ergs on 06T08:36 UT. This indicates that the AR reaches to higher non-potential state by this time and an X2.2 flare was seen peaking at 06T09:10 UT. The total magnetic energy is less by $2.8\times10^{32}$ by 06T10:36 UT compared to the earlier panel indicating energy during the flare. However, it increased to $25.9\times10^{32}$ ergs by 06T11:24 UT showing energy storage for the X9.3 flare which occurred commencing from 06T11:53 UT. Since the free energy is above $5.6\times10^{32}$ ergs at any time, an M or X class flare associated with CME is expected depending on the other favorable non-potential conditions as discussed in the section 3.1. We can compare these energy estimations to that derived from energy injection method (equation~\ref{eq_dedt}) by time integration as $E_{acc}=\int_0^T (dE/dt)\,dt$. As can be seen in Table~\ref{tab1}, the $E$, $E_{acc}$ agrees each other by orders of magnitude except the fact that $E$ corresponds to the activity events indicating storing/releasing phase unlike the monotonous accumulation in the later. 

\begin{table*}[!ht]
	\centering
	\caption{Quantitative estimates of helicity, energy, and flux rope twist. }
	\begin{tabular}{ccccccc}
		\hline 
		time [UT]& $H_{acc} [10^{43}Mx^2]$ & $E_{acc} [10^{32} erg]$ & $H_R [10^{43}Mx^2]$ & $E_p [10^{32}erg]$  & $E [10^{32}erg]$ & $<T_w >[turns]$  \\
		\hline
		2017-09-04T18:00  & -2.5 & 8.8   & -4.66&12.4 &18.3 &-1.22 \\ 
		2017-09-05T12:00 & -4.4 & 14.1  & -5.33&14.9 &20.9 &-0.37\\
		2017-09-06T00:12 & -5.7 & 17.3  & -5.73&16.8 &23.2 &-0.34 \\
		2017-09-06T08:36 & -6.4 & 19.1  & -6.83&18.2 &26.9 &-0.87\\
		2017-09-06T10:36 & -6.6 & 19.9  & -6.06&18.5 &24.1 &-0.77 \\
		2017-09-06T11:24 & -6.8 & 20.5  & -6.72&18.7 &25.9 &-0.93 \\
		\hline 
	\end{tabular} 
	\label{tab1}
	
\end{table*}

From the helicity injection method discussed in section 3.1.4, the coronal magnetic helicity is estimated as $H_{acc}=\int_0^T(dH/dt)\,dt$. Alternatively, this can also be computed from volumetric distribution of magnetic field above the AR, which is generally a model like force-free extrapolation \citep[for more details on different methods, refer to][]{Valori2016}. To compare the coronal helicity budget, we calculate relative magnetic helicity \citep{Berger1984} from the 3D modeled field as given by 
\begin{equation}
H_{R}=\int_V \left(\mathbf{A}+\mathbf{A_p}\right)\cdot\left(\mathbf{B}-\mathbf{B}_p\right)dV
\end{equation}
Here the reference field is potential field $\mathbf{B}_p$ and $\mathbf{A}_p$ is the corresponding vector potential. The main assumption involved is that the reference field should have the same normal field component as that of the real magnetic field on the boundaries. We construct the vector potentials with the formalism given in \citet{Devore2000} under the Coulomb gauge condition as 
\begin{equation}
\mathbf{A}_p(x,y,z)=\nabla\times \hat{z}\int_z^\infty dz' \phi(x,y,z)
\end{equation}
where the scalar function obeys the Laplace equation $\nabla^2\phi=0$ recovering the potential field such that $\nabla\times\mathbf{A}_p=-\nabla\phi$. Using the Green's function for Laplace equation as integral kernel, 
\begin{equation}
\phi(x,y,z)=\frac{1}{2\pi}\int\int dx'dy' \frac{B_z(x,y,z=0)}{\left[(x-x')^2+(y-y')^2+z^2\right]^{1/2}}
\end{equation}
This result is used to calculate the vector potential $\mathbf{A}_p$ at $z=0$. These boundary values are used to compute the $\mathbf{A}$ for the actual magnetic field $\mathbf{B}$, by integration
\begin{equation}
\mathbf{A}(x,y,z)=\mathbf{A}_p(x,y,0)-\hat{z}\times\int_0^z dz' \mathbf{B}(x,y,z')
\end{equation}
The values of $H_R$ and $H_{acc}$ are tabled in Table~\ref{tab1}. They match each other in orders of magnitude. A point to be noted is that the information of helicity taken away by intermittent CMEs is not included in $H_{acc}$ and show monotonous increase in time whereas the $H_R$ decreases after X2.2 flare (10:36 UT) and then increases (11:24 UT) towards  X9.3 flares on September 6. Moreover, $H_R$ is consistent with the $E$ for being the fact that they are related each other proportionally. 

Given 3D field distribution, one can calculate the twist number for each field line \citep{berger2006,inoue2011, LiuRui2016}
\begin{equation} \label{Eq-twist}
T_w = \int\limits_{L} \frac{\mu_0 \mathbf{J}_{||}}{4\pi B} ~\rmd l
= \int\limits_{L} \frac{\nabla \times \mathbf{B} \centerdot \mathbf{B}}{4\pi B^2} ~\rmd l
\end{equation} 
Here the twist is related to the parallel electric current given by ${{\mathbf{J}}_{||}}=\frac{\mathbf{J}\centerdot \mathbf{B}}{\left| B \right|}$ and the line integral is along the selected magnetic field line of length $L$.  The average twist number then is given by

\begin{equation}
<T_w>=\frac{\Sigma_i \Phi_i T_{w,i}}{\Sigma_i \Phi_i}
\end{equation}
where $\Phi_i$ is the magnetic flux of flux tube i. These are estimated in the flux rope cross section as shown in the last column of Figure~\ref{Fig_qf}. The flux rope boundary is assumed as the QSLs of the large Q-values, within which the field lines are twisted. The values of $<T_w>$ are furnished in Table~\ref{tab1}. The $T_w$ values in the flux rope at 04T18:00UT ranges to 2 turns (negative is for left hand helicity) but the $<T_w>$ becomes 1.2 turns indicating favorable condition for kink-instability as also reported by \citet{YangS2017}. In the 05T12:00 UT and 06T00:12 UT time shots, the $<T_w>$ is small for the fact that the field lines are in sheared arcade form. $T_w$ has maximum values of 1.72 in the 06T08:36 UT panel, but the average value comes down to 0.87 turns. In the later times, the $<T_w>$ recovers to 0.99 turns before the large X9.3 flare at 11:53 UT. Further, the $<T_w>$ values correspond to $H_{rel}$ values at different epochs because the later is the measure of twist, linking, and shear which mostly come from the core part. The $<T_w>$ values represent the physical structure of magnetic field derived from the simultaneous observations, thus are not in agreement with the $H_{acc}$ which in itself has no information of the loss of helicity by flux rope eruption i.e., flux rope present or not. Helicity injection builds the coronal structure of flux rope, expecting the buildup of the flux rope twist with the coronal helicity accumulation as reported in the modeling analysis of \citet{Guoy2013}. However, this is possible as long as no eruption occurs in the observation time window. In any AR since its emergence, helicity builds to form structures like flux ropes which erupt after some critical value. A persistent injection of helicity would builds such structures over a time scale of tens of hours after every eruption successively and has been the subject of recent studies \citep{vemareddy2017b}. Altogether, our estimated parameters are evidently shown to change with the flux rope presence or its eruption. This modeling analysis quantitatively captures the most of the theoretical aspects as seen with the coronal activity.     

From the $T_w$ computation, we can also estimate the self-helicity of the flux rope as $H_s=\Sigma_i ^N T_{w,i}\Phi_i^2$, with the summation running over N flux tubes. We use the flux rope cross section defined by the closed-QSLs in the boundary. The values vary between $0.3-1.0\times 10^{40}$ Mx$^2$ at different time snapshots. This is consistent with the previous findings of \citet{Guoy2013,Guoy2017b} that the self-helicity becomes negligible for large N.  

\begin{figure*}[!htp]
	\centering
	\includegraphics[width=.9\textwidth,clip=]{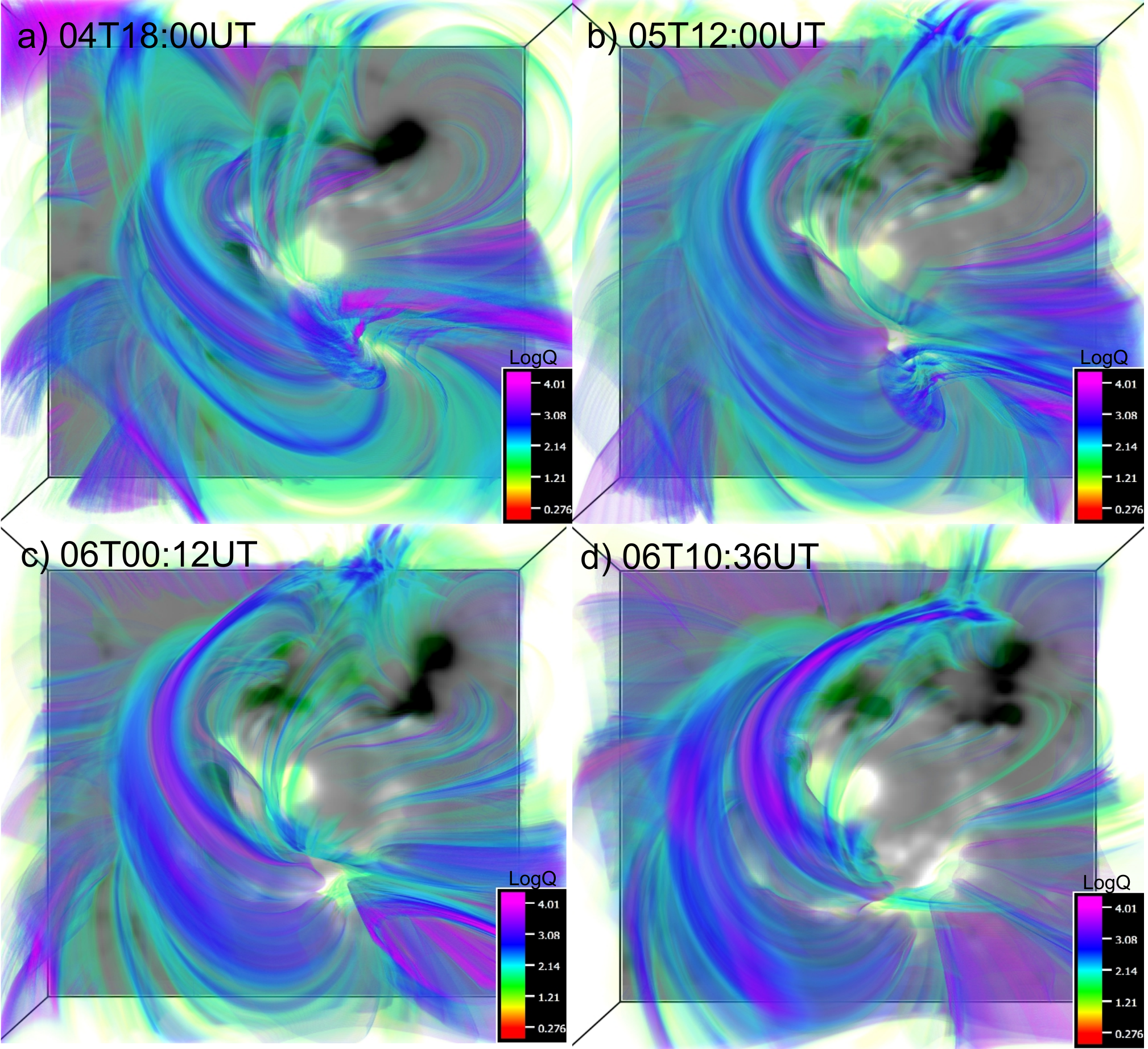}
	\caption{ 3D distribution of $LogQ$ for the NLFFF magnetic structure in AR 12673. The background is $Bz$ image. Notedly, the Q-distribution outlines the sigmoid structure with the field lines at the core of the flux rope, the overlying potential arcade, and 2J field lines. Higher values of $LogQ$ are located in the core part of the sigmoid. An online only animation for 04T18:00\,UT shot describes the relationship of the Q-distribution with variable opacity and the field lines. }	\label{Fig_q3d}
\end{figure*}

\subsection{QSLs and magnetic connectivity domains}
Quasi separatrix layers (QSLs) are the regions of the magnetic volume where the field line connectivity experiences dramatic but continuous variations \citep{demoulin1996}.  The locations of QSLs are determined by computing the squashing factor (Q; \citealt{titov2002}). From the 3D PF and NLFFF, we calculate Q using the code developed by \citet{LiuRui2016} according to the formalism given in \citet{Pariat2012}. In Figure~\ref{Fig_q3d}, the 3D-rendering of $LogQ$ is shown for different time shots of NLFFF structure. There are large Q values upto 15 orders, but a scaling of $0<LogQ<4$ is applied in order for a better visualization of all range. As is obvious from the panels, the Q-distribution outlines the sigmoid structure with the field lines in the core of the flux rope, the overlying potential arcade, and 2J field lines, similar to the QSL analysis of \citet{Tassev2017} for the \citet{titov1999} flux rope. Higher values of Q are located in the core part of the sigmoid above the main PIL. Online only available animation demonstrates the relationship of the Q-distribution with variable opacity and the field lines. QSLs (in blue) above the main PIL majorly separates the quasi-connectivity domains from P1 and N1 polarities.  The QSLs of large Q-values are likely places for reconnection, in which the line-of-sight integrated emission resembles the given shape for the X-ray or EUV sigmoid. However, all QSLs of higer Q-values are not associated with the intense volume currents as described in \citep{savcheva2012b,Guoy2013,YangKai2015}.    

\begin{figure*}[!ht]
	\centering
	\includegraphics[width=.99\textwidth,clip=]{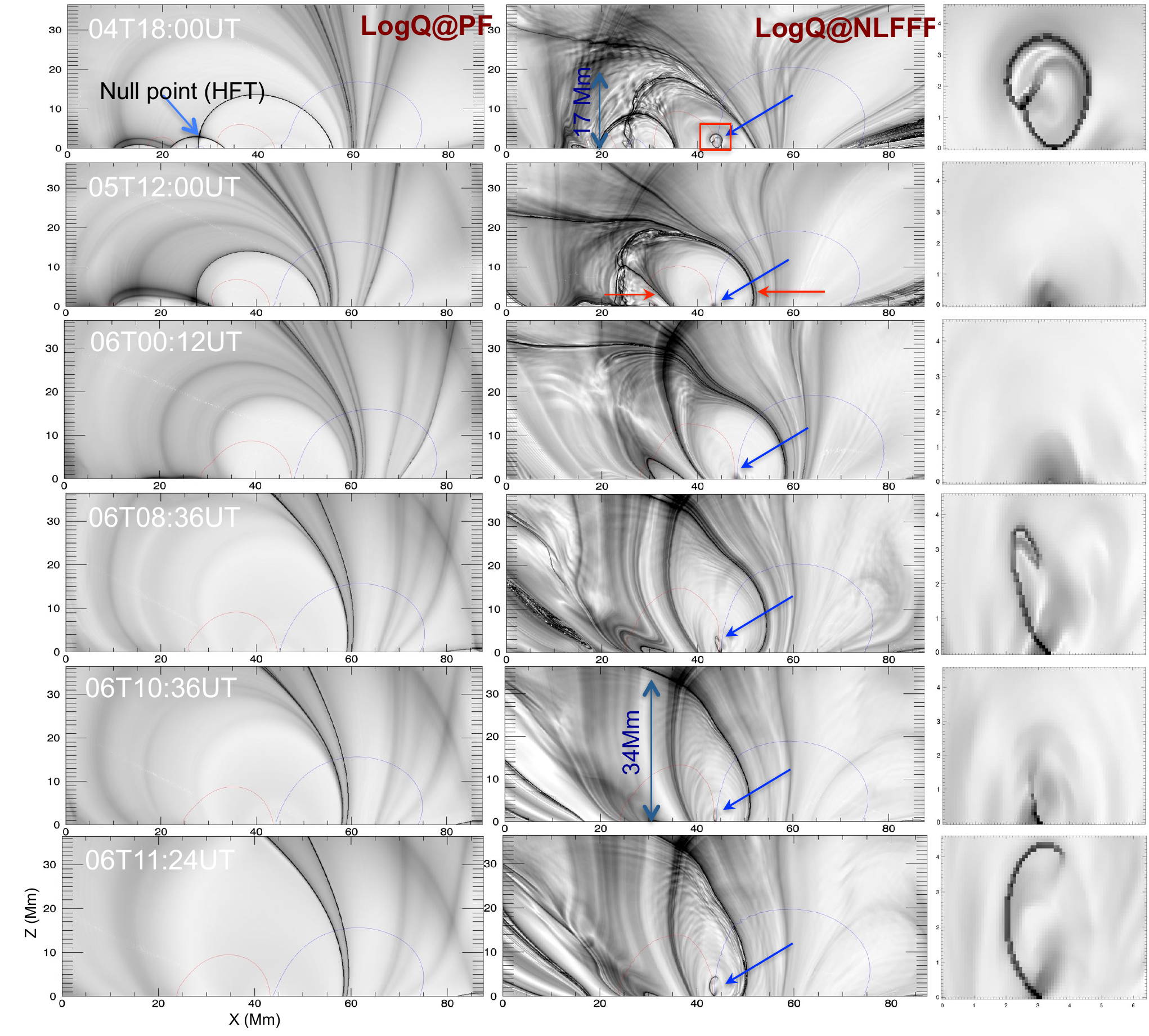}
	\caption{Distribution of $Q$ in a vertical slice (xz plane, along magenta line in Figure~\ref{Fig_core_fld}) placed across the flux rope. {\bf first Column:} LogQ computed from potential field approximation. QSL of two domains intersect at null point in the corona (see Figure~\ref{Fig_null_fr}) in the 04T18:00 UT panels and disappears in later time panels.  {\bf second column:} LogQ computed from NLFFF model. Thin blue/red curves represent $B_z$ contours at $\pm100$\,G in the slice and their adjoining position on the photosphere (z=0) traces the main PIL. The coronal volume above the main PIL is surrounded by magnetic domains of highly sheared arcade enclosed in a less sheared potential like field. The inner core of the sheared arcade is twisted flux rope extended upto 5 Mm and is pointed by blue arrows.  With the continued shear motion of magnetic patches about the PIL, the existing arcade becomes highly stressed (red arrows) resulting in its increased coronal height from 17 Mm to 34 Mm.  {\bf third column:} QSL of the flux rope cross-section enclosed in the red rectangle.  }
	\label{Fig_qf}
\end{figure*}

Further, the Q (in logarithmic scale) in a vertical slice placed across the flux rope (see Figure~\ref{Fig_ext_j2d}) is plotted in Figure~\ref{Fig_qf}. $B_z$ contours at $\pm$80G are also shown with thin blue/red curves. The joining locations of these contours at the photosphere (z=0) corresponds to the main PIL and is surrounded by magnetic domains of sheared arcade enclosed in a less sheared potential field. The important QSLs are identified by large values of Q i.e.,  the black lanes (or patches) in our negated maps. Owing to the presence of stressed field configuration, there is clear difference in the domains of sheared arcade in both the PF and NLFFF, especially the candle-flame or inverse tear-drop like shape in the later. The inner core of the sheared arcade is highly twisted manifesting a flux rope extended upto 6\,Mm (pointed by blue arrow). 

To be more clear, the Q maps for flux rope cross section (red rectangle) is displayed in third column of the same figure. The large values of Q in the 04T18:00 UT panel are located in the flux rope border which differentiates the domains of flux rope with twisted field lines and the surrounding sheared arcade, similar to the studies of \citet{Guoy2013,Jiezhao2014, LiuRui2016}. However, this flux rope structure is not clear with large QSLs in the border for the 05T12:00 UT, 06T00:12 UT panels probably because flux rope was not formed, after the first phase of flaring activity from 04T18:00 UT, from the sheared arcade. This is consistent with the inference of \citet{YangKai2016} who found that the closed-QSL of flux rope becomes smaller as a consequence of the flare. In the 06T08:36UT panel, these QSLs appear to be candle flame shape with less width compared to 04T18:00 UT case. And this structure is diffused in 06T10:36 UT panel, probably for the fact that the flux rope might partially erupted during X2.2 flare at 06T08:57 UT. The 06T11:24 UT panel clearly shows the semi-closed QSL implying a developed state of the flux rope before X9.3 flare at 11:53 UT. We should point from these QSL analysis that the flux rope has an HFT topology from 06T08:36 UT onwards as the QSL legs cross each other below as also revealed by the numerical study of this AR by \citet{JiangChaowei2018}. Differently, the topology would be BPSS in 04T18:00 UT panel as the QSLs touch the photosphere tangentially.   

Further, there are other QSLs associated to large scale magnetic structure enclosing the flux rope. These are also self-closed QSLs differentiating less sheared field lines over the flux rope and the potential ones in even higher height. Importantly, there exists intersecting self-closed QSLs in the 04T18:00 UT. The intersection is at the null point above the positive polarity on the east side of the flux rope as displayed in Figure~\ref{Fig_null_fr}. Thus the spine field lines extends in the P1, P2 over the flux rope. The reconnection in the null point helps to reduce the overlying field,  as in a breakout reconnection scenario, enabling the flux rope eruption \citep{WuST2005,Torok2009}. As the positive polarity moves southward, these intersecting QSLs become separated after 05T12:00 UT panels.  

It is important to note the height of the large scale QSLs over the time evolution. They appears at increasing height from 04T18:00 UT panel to 06T10:36 UT panel. In the presence of persistent slow shear motions of the magnetic patches (N1 and P1) at the photosphere (Figure~\ref{Fig_vel}), the field lines are increasingly stressed. This results in the decreased distance of the QSL legs rooted in the photosphere on either side of the PIL. The configuration still remains in equilibrium by increasing the extent of sheared arcade height from 18 Mm on 04T18:00UT to 34 Mm on 06T10:36 UT. We believed that this highly energized sheared arcade system is critically stable and is prone to erupt to a small perturbation like kink-instability of the flux rope in the inner core \citep{YangS2017}.       

\begin{figure*}[!htp]
	\centering
	\includegraphics[width=.98\textwidth,clip=]{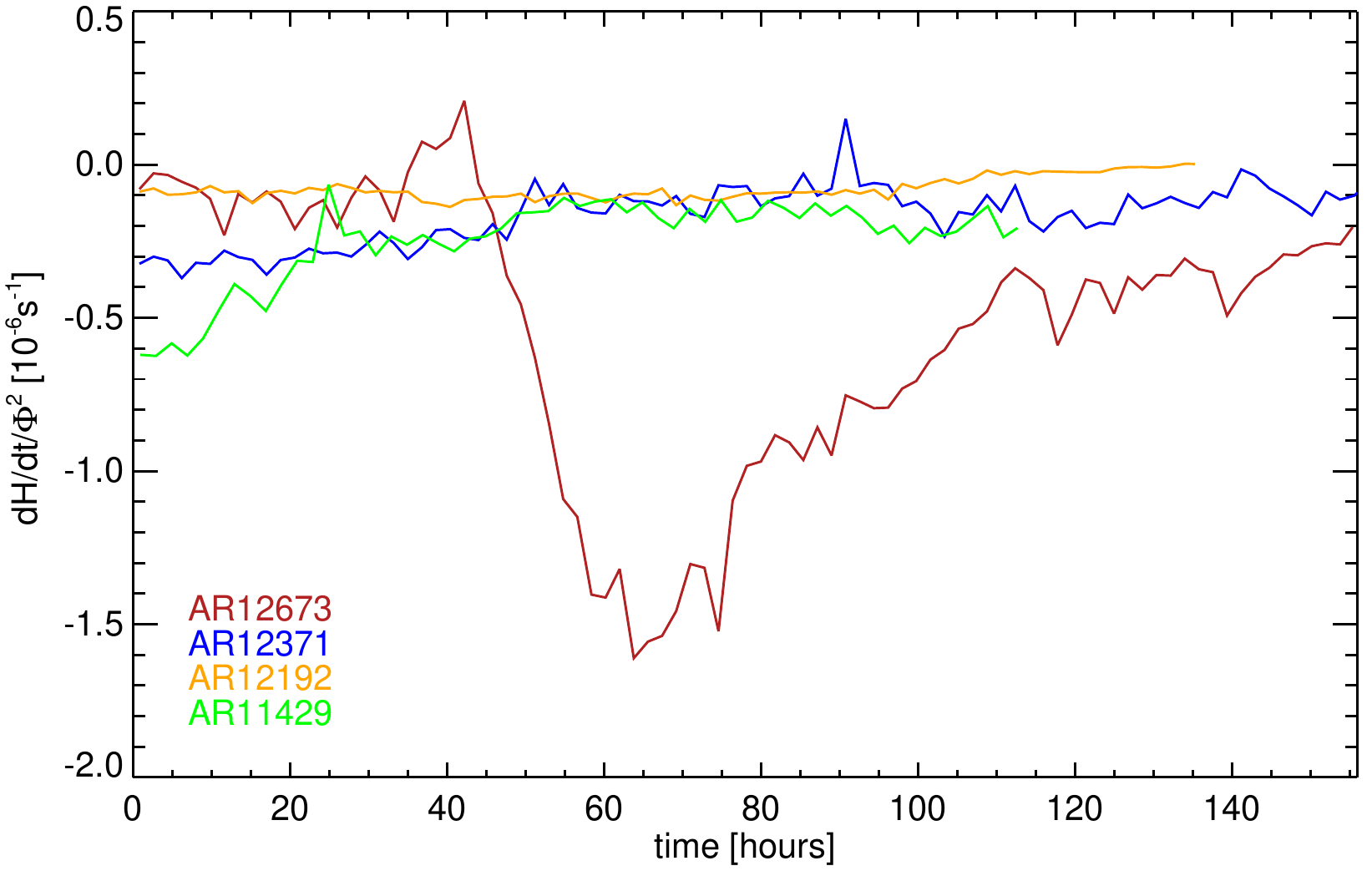}
	\caption{Comparison of normalized helicity flux injection $dH/dt/\Phi^2$ in 4 flare/CME producing ARs. Starting times are 2017-08-31T19:12 UT,   2015-06-19T00:00 UT, 2014-10-21T00:00 UT, 2012-03-06T00:00 UT , for AR 12673, 12371, 12192, 11429 respectively. Normalized helicity flux injection in AR12673 is comparatively high with major injection phase co-temporal with rapid flux emergence. The mean values in the given time windows of these ARs are noted as $-0.52\times10^{-6}$ s$^{-1}$, $-0.19\times10^{-6}$ s$^{-1}$, $-0.08\times10^{-6}$s$^{-1}$, $-0.25\times10^{-6}$s$^{-1}$ respectively. }
	\label{Fig_norm_dhdt}
\end{figure*}
\section{Summary and Discussion}
\label{summ}
Using continuous time-series SDO observations, we studied the long term-evolution of flare-prolific AR 12673 right from its emergence on the visible photosphere. The AR emergence occurred with a couple of dipoles in the vicinity of pre-existing positive polarity sunspot whose interaction by shear/proper motions builds into compact $\delta$-AR complex with curved SPIL. This main SPIL is seen with persistent strong shear and converging flows of opposite polarities on either side. A major helicity injection occurs during rapid flux emergence consistent with the very fast flux emergence cases \citep{SunX2017}.  

In order to have more insight on helicity flux input, we compared the normalized helicity flux ($\frac{1}{\Phi^2}\frac{dH}{dt}$) injection in different ARs in Figure~\ref{Fig_norm_dhdt}. The value is a measure of non-potentiality per unit flux per unit time and indicates how fast the AR accumulates energy and helicity. As is clear from the plot, this parameter evolves at a  higher rate by a factor of 3 in AR 12673 compared to other ARs. The mean values of $\frac{1}{\Phi^2}\frac{dH}{dt}$ in the given time windows of these ARs are noted as $-0.52\times10^{-6}$ s$^{-1}$, $-0.19\times10^{-6}$ s$^{-1}$, $-0.08\times10^{-6}$ s$^{-1}$, $-0.25\times10^{-6}$ s$^{-1}$ for AR 12673, 12371, 12192, 11429 respectively. The AR 12192 is a flare-prolific region without CMEs \citep{sunx2015} in contrast to the rest of the ARs \citep{vemareddy2017b} and has small injection value per flux tube, Interestingly, this value in our AR is quite stronger by a factor of 2 than the strong CME-prolific ARs 12371, AR11429 which is suggested to be key parameter for generating severe space-weather events as the case in AR12673.    

While this helicity flux builds up the sigmoid by September 4, the helicity injection by the continued shear and converging motions in the later evolution contributes to sigmoid sustenance and its core field twist as a manifestation of the flux rope which erupts after exceeding critical value of twist. Moreover, the total length of SPIL segments correlates with the non-neutralized current ($|DC/RC|$) and maintains a higher value in both the polarity regions in the AR. This higher value of non-neutralized currents is a signature of strong non-potentiality and eruptive capability of the AR according to the flux rope models  \citep{zakharov1986,Torok2014}, and is suggested to be a proxy assessing the ability of ARs to produce major eruptions \citep{YangLiu2017,vemareddy2017d, Vemareddy2019}. 

Corresponding to the photospheric magnetic field evolution, we also studied the magnetic configuration by modeling the coronal field. The modelled magnetic field qualitatively reproduces the sigmoidal structure capturing major features like twisted core flux as flux rope, and hook-shaped parts connecting P1 and N3 polarity regions. Topological study indicates that the AR consists of QSLs in the surroundings of the flux rope in the core and those in the large structure surrounding the sheared arcade. The flux rope was likely having BPSS topology during the emergence phase (04T18:00 UT) of the AR whereas it was HFT in later time on September 6. However, the twist number of the field lines in the flux rope have more than 1.2 turns (with $<T_w>\ge1$ turn) indicating a possible kink-unstable nature of the twisted flux in the core. In addition, the magnetic structure reveals a null point topology with spine field lines overlying the flux rope. Therefore, both kink-instability and null-point reconnection in the overlying field would have played role in triggering the first phase of the activity from September 4 (Figure~\ref{Fig_met}). In the second phase of the activity on September 6, although the AR was critically stable, kink-instability and/or null point reconnection below the flux rope might play the role of triggering the eruptions, as also revealed by numerical studies of \citet{JiangChaowei2018}. 

Further, the twist number of the flux rope, the QSL structure surrounding the flux rope and coronal helicity and energy budgets are shown to change with the presence of flux rope and its eruption. This implies that the the energy and helicity injections from the bottom boundary help build the essential physical condition for flare/CME occurrence in the long term evolution. QSL study shows that the sheared arcade is stressed to a critically stable state and its coronal height becomes doubled from September 4 to 6. In addition to the many distinguished non-potential characteristics, the critically stable state from the mid of September 6 is a crucial factor to explain the recurrent eruptive nature of this AR. Once the magnetic system is critically stable, there are basically three alternatives for triggering solar eruptions viz., internal tether cutting, external tether-cutting, and MHD instability or loss of equilibrium such as kink-instability and/or torus-instability \citep{linj2000,antiochos1999,torok2005}. The unique non-potential characteristics found in this study are in agreement with previous studies and interpretations. Studies of this AR 12673 by \citet{YangS2017} showed that the flux rope eruption is triggered by kink-instability because of the exceeding critical twist. They had proposed a block-induced complex structure for a flare-productive AR. Similarly, \citet{YanXL2018} suggested the sunspot rotation and shearing motion played an important role in the buildup of free-energy and the formation of flux ropes in the corona which produces solar flares and CMEs. We believe that the long term evolution of the AR has characteristic clues for the nature and strength of the activity, which by studying different AR cases can help constrain the space-weather prediction models.  

\acknowledgements SDO is a mission of NASA's Living With a Star Program. I thank the anonymous referee for insightful comments and suggestions that undoubtedly improved the presentation of the results. P.V is supported by an INSPIRE grant under AORC scheme of Department of Science and Technology.  The NLFFF code is developed by Dr. T. Wiegelmann of Max Planck Institute for Solar System. 3D rendering is due to VAPOR (\url{www.vapor.ucar.edu}) software. We acknowledge an extensive usage of the multi-node, multi-processor high performance computing facility at Indian Institute of Astrophysics. 

\bibliographystyle{apj} 
%\bibliography{../ref_lib}    

%%%%%%%%%%%%%%%%%%%%%%%%%%%%%
%
%%	  FIGURES						%%
%
%%%%%%%%%%%%%%%%%%%%%%%%%%%%%

\end{document}